\documentclass[prd,twocolumn,a4paper,superscriptaddress,floatfix]{revtex4}
\usepackage{graphicx}
\usepackage{amssymb}
\usepackage{amsfonts}
\usepackage{amsmath}
\usepackage{color}
\usepackage{ulem}
\begin{document}

\newcommand{\be}{\begin{equation}}
\newcommand{\ee}{\end{equation}}
\newcommand{\bq}{\begin{eqnarray}}
\newcommand{\eq}{\end{eqnarray}}
\newcommand{\bsq}{\begin{subequations}}
\newcommand{\esq}{\end{subequations}}
\newcommand{\bc}{\begin{center}}
\newcommand{\ec}{\end{center}}
\newcommand {\R}{{\mathcal R}}
\newcommand{\al}{\alpha}
\newcommand\lsim{\mathrel{\rlap{\lower4pt\hbox{\hskip1pt$\sim$}} \raise1pt\hbox{$<$}}}
\newcommand\gsim{\mathrel{\rlap{\lower4pt\hbox{\hskip1pt$\sim$}} \raise1pt\hbox{$>$}}}
\newcommand{\mdj}[1]{\textcolor{green}{ - #1 -}}
\newcommand{\rcq}[1]{\textcolor{cyan}{ - #1 -}}

\title{Soft Pomeron in Holographic QCD}

\author{Alfonso Ballon-Bayona}
\affiliation{Centro de F\'{\i}sica do Porto e Departamento de F\'{\i}sica e Astronomia da Faculdade de Ci\^encias da Universidade do Porto, Rua do Campo Alegre 687, 4169-007 Porto, Portugal}
\author{Robert Carcass\'{e}s Quevedo}
\affiliation{Centro de F\'{\i}sica do Porto e Departamento de F\'{\i}sica e Astronomia da Faculdade de Ci\^encias da Universidade do Porto, Rua do Campo Alegre 687, 4169-007 Porto, Portugal}
\author{Miguel S. Costa}
\affiliation{Centro de F\'{\i}sica do Porto e Departamento de F\'{\i}sica e Astronomia da Faculdade de Ci\^encias da Universidade do Porto, Rua do Campo Alegre 687, 4169-007 Porto, Portugal}
\affiliation{Theory Division, Department of Physics, CERN,CH-1211 Gen\`eve 23, Switzerland}
\author{Marko Djuri\'c}
\affiliation{Centro de F\'{\i}sica do Porto e Departamento de F\'{\i}sica e Astronomia da Faculdade de Ci\^encias da Universidade do Porto, Rua do Campo Alegre 687, 4169-007 Porto, Portugal}


\begin{abstract}
We study the graviton Regge trajectory in Holographic QCD as a model for  high energy scattering processes dominated by
soft pomeron exchange. This is done by considering spin $J$ fields from the closed string sector that are dual to glueball
states of even spin and parity. In particular, we construct a model that governs the analytic 
continuation of the spin $J$ field equation to the region of real $J<2$, which includes the scattering domain of negative Maldelstam variable $t$.
The model leads to approximately linear Regge trajectories and is compatible with the measured values 
of 1.08 for the intercept and $0.25$ ${\rm GeV}^{-2}$ for the slope of the soft pomeron.
The intercept of the secondary pomeron 
trajectory is in the same region of the  subleading  trajectories, made of mesons, proposed by Donnachie and Landshoff,
and should therefore be taken into account.
\end{abstract}
\maketitle

\section{\label{sec:Int}Introduction}

The Pomeron plays a crucial role in QCD Regge kinematics, for 
processes dominated by  exchange of the vacuum quantum numbers.
This includes elastic scattering of soft  states at high energies and low momentum transfer.
The corresponding amplitude  exhibits  a universal behavior explained within  Regge theory \cite{Donnachie:1992ny},
\begin{equation}
{\cal A}(s,t) \approx \beta(t) \,s^{\alpha(t)}\,,\ \ \ \ \alpha(t)= 1.08 + 0.25\,t\,,
\label{eq:SoftPomeron}
\end{equation}
in ${\rm GeV}$ units and for some function $\beta(t)$ that depends on the scattered states. 
A precise computation of the values of the intercept ($\alpha_0=1.08$) and slope ($\alpha'=0.25 \ {\rm GeV}^{-2}$)
is beyond our current understanding of QCD, since long-range strong interaction effects are important. 

The gauge-gravity duality is a new tool to unveil QCD strongly coupled physics \cite{AdSCFT}. 
In particular, the Pomeron is conjectured to be
the graviton Regge trajectory of 
the dual string theory \cite{Brower1}. This fact has been explored in diffractive processes 
dominated by Pomeron exchange, like low-$x$ deep inelastic scattering (DIS) \cite{Saturation,Levin:2010gc,Brower:2010wf},
deeply virtual Compton scattering \cite{Costa:2012fw}, vector meson production \cite{Costa:2013uia} 
and double diffractive Higgs production \cite{Brower:2012mk}. 

Consider for instance the case of low-$x$ DIS.  One observes a rise of the intercept $j_0$ from 1.1 to 1.4 as $Q$ grows, where Q is the
momentum scale of the photon probe.
The conventional approach is to start from the perturbative BFKL hard pomeron \cite{BFKL}, which still exhibits conformal symmetry.
Introducing a cut-off, one explains the observed rise of the structures functions and even the running of the intercept, provided the cut
$Q^2>4\ {\rm GeV}^{2}$ is imposed in the kinematics \cite{Kowalski:2010ue}. 
However,  dual models that also start from a conformal limit and introduce a hard wall cut off in AdS space,
give even better fits to data, without imposing any restriction in the kinematics \cite{Brower:2010wf}. 
This is a strong motivation in favor of treating soft  pomeron physics using the gauge/gravity duality.

This letter builds a soft-pomeron phenomenology  in Holographic QCD. More concretely,
we show the  Regge theory for spin $J$  exchanges in the dual geometry leads to the 
behavior (\ref{eq:SoftPomeron}) for the  amplitude between soft  probes. 

\section{\label{sec:AdS/QCD}Holographic QCD model}

We will consider the Holographic QCD model proposed in the works  \cite{Gursoy:2007cb,Gursoy:2007er,Gursoy:2010fj} based on gravity plus a dilaton field. We shall be working in the string frame because the Regge trajectory we are interested in is made of fundamental closed string states.
As usual, the scalar field $\Phi=\Phi(z)$ and the dual geometry has metric
\begin{equation}
ds^2=  g_{ab} dx^a dx^b =e^{2A(z)} \left( dz^2 +\eta_{\alpha\beta}dx^\alpha dx^\beta\right)\,,
\label{eq:metric}
\end{equation}
where $\eta_{\alpha\beta}$ is the Minkowski boundary metric.
In the string frame the corresponding action is
\begin{equation}
S= \frac{1}{2\kappa^2} \int d^5 x \sqrt{-g} \,e^{-2\Phi}  \left[Ê R + 4 \left( \partial \Phi  \right)^2  + V \right] \, ,
\label{eq:Einstein-DilatonAction}
\end{equation}
with
\begin{equation}
V = e^{-\frac43 \Phi} \left [ \frac{64}{27} \,W^2 - \frac43 \left (\frac{d W}{d\Phi} \right )^2 \right ].
\end{equation}
The  field  $\Phi$ is the dilaton without the zero mode that is absorbed in the gravitational coupling $\kappa$.
The field equations arising from (\ref{eq:Einstein-DilatonAction}) take the form
\begin{align}
R_{ab} + 2 \nabla_a  \nabla_b \Phi   - \frac14 \frac{d V}{d \Phi} g_{ab}  =& 0 \, , \cr
  2 \nabla^2 \Phi  - 4 (\nabla \Phi)^2 + V  + \frac34 \frac{d V}{d \Phi} =& 0 \,.
  \label{eq:Einstein-Dilaton}
\end{align}
The superpotential $W(\Phi)$ is fixed  phenomenologically  by demanding that the model reproduces basic QCD data, such as beta function, heavy quark/anti-quark
linear potential  and glueball spectrum. In this work we take  the Background I of \cite{Gursoy:2007cb,Gursoy:2007er} where 
\begin{equation}
W= \frac{9}{4L} \left (1 + \frac23 b_0 \lambda \right )^{\frac{2}{3}} 
\!\left [ 1 + \frac{(2 b_0^2 + 3 b_1 ) \log (1+ \lambda^2 ) }{18 a} \right ]^{\frac{4a}{3}} \!, 
\label{eq:Superpotential}
\end{equation}
$\lambda = e^{\Phi}$ and the length scale $L$ fixes the units. 

The  't Hooft coupling of the dual Yang-Mills theory $\bar \lambda$ is fixed by $\lambda$ up to a multiplicative constant, i.e. $\bar \lambda = c_0 \lambda$. For the model considered in this work the constants in (\ref{eq:Superpotential})
are given by
\begin{equation}
b_0 = 4.2 \,, \quad 
\frac{b_1}{b_0^2} = \frac{51}{121} \, , \quad
a = \frac{3}{16} \, . \label{eq:ModelParameters}
\end{equation}
The model has an additional integration constant that can be related to $\Lambda_{QCD}$, via 
the identification of the energy scale and warp factor, $\log E = A(z) - \frac23 \Phi(z)$.
As shown in \cite{Gursoy:2007cb,Gursoy:2007er}, the UV behavior of the superpotential (\ref{eq:Superpotential}) leads to a beta-function 
\begin{equation}
\beta = \frac{d\lambda}{d\log{E}} = - b_0 \lambda^2 - b_1 \lambda^3 + \dots \,. 
\end{equation}
This is consistent with the two-loop perturbative beta function in large-N Yang-Mills
\begin{equation}
\bar \beta = - \bar b_0 \bar \lambda^2 - \bar b_1 \bar \lambda^3 
\, , \quad \bar b_0 =\frac23 \frac{11}{(4 \pi)^2}  \, , \quad \frac{\bar b_1}{\bar b_0^2} = \frac{51}{121}  \,. 
\label{eq:beta_func}
\end{equation}
if we take $c_0=b_0/\bar b_0$. This fixes the second parameter in (\ref{eq:ModelParameters}). 
The others parameters are fixed by the IR constraints coming from confinement and asymptotic linear glueball spectrum and lattice QCD. 

\begin{figure}[t!]
\begin{center}
\includegraphics[height=7cm]{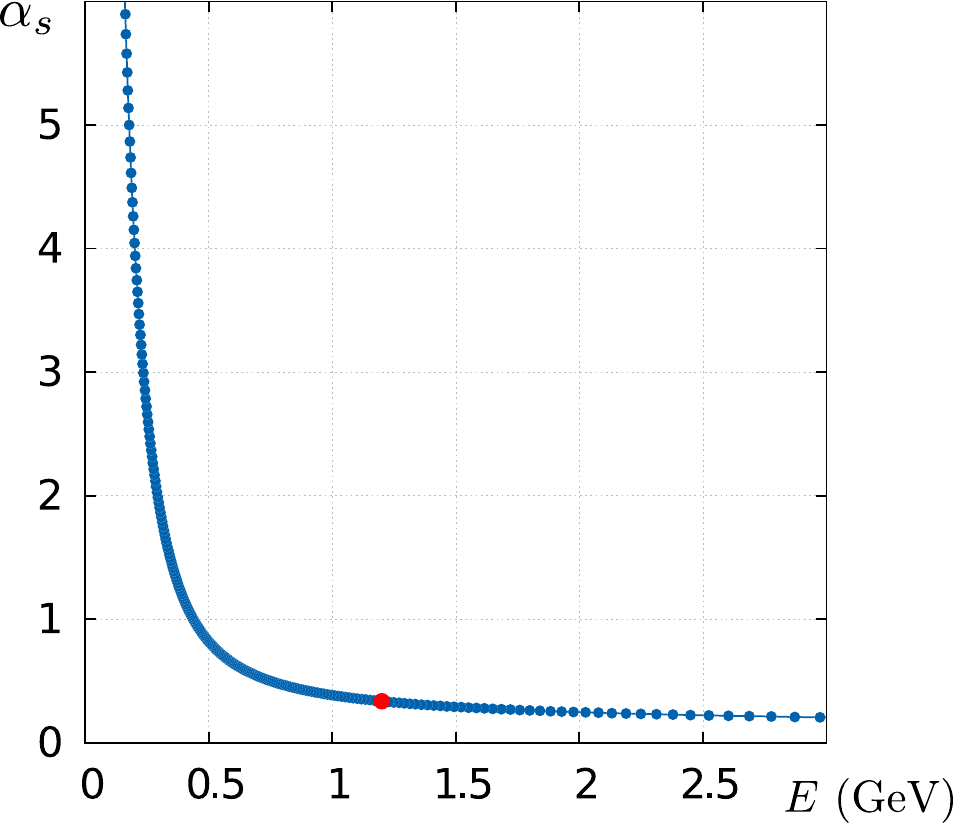}
\end{center}
\vspace{-0.5cm}
\caption{Running coupling $\alpha_s$ vs. energy scale. The red point is $\alpha_s(1.2\, \text{GeV})=0.34$.}
\label{fig:running_coupling}
\end{figure}

Given that all of the parameters are already fixed at  this point, one may ask how the field theory coupling runs with energy.
Setting $N_c=3$ the QCD running coupling can be identified with $\alpha_s=\bar{\lambda}/(12\pi)$. Figure \ref{fig:running_coupling} shows how $\alpha_s$ runs with the energy scale in the model, giving $0.34$ for the value for $E=1.2\, \text{GeV}$, which is very close to the experimental value 0.35.

We can recover the conformal limit by considering the parameters $b_0 = b_1 =0$. Then we can set  $\Phi =0$, the superpotential becomes the cosmological constant $-12/L^2$, and the metric becomes
that of AdS space, i.e. $A(z) = \ln (L/z)$.

\section{\label{sec:ReggeTheory}Pomeron in Holographic QCD}
 
\subsection{\label{sec:Graviton}Graviton}

Since we are interested in the graviton Regge trajectory let us start by considering perturbations to the background in the string frame.
We shall write the metric and dilaton,  respectively, as
\begin{equation}
g_{ab} + h_{ab}\,,\quad \quad \quad
\Phi + \varphi\,.
\end{equation}
It is then a mechanical computation to obtain from (\ref{eq:Einstein-Dilaton}) the linearised equations of motion 
for the perturbations $h_{ab}$ and $\varphi$,
\begin{align}
\nabla^2 h_{ab}  -2 \nabla_{(a} \nabla^{c} h_{b)c}  + \nabla_a\nabla_b h  + 2 R_{acbd} h^{cd} 
\nonumber\\
+ 4 \nabla^c\nabla_{(a}  \Phi  \,h_{b)c}+2 \nabla^c\Phi \left(   2 \nabla_{(a} h_{b)c} - \nabla_c h_{ab}\right) 
\nonumber\\
-4 \nabla_a\nabla_b\, \varphi +\frac{1}{2}\,g_{ab} V''(\Phi) \,\varphi=0\,,
\label{eq:flutuations}
\\
\nabla^2 \varphi +\frac{1}{2} V'(\Phi) \,\varphi +\frac{3}{8} V''(\Phi) \,\varphi
\nonumber\\
-4\nabla\varphi \cdot \nabla\Phi - \frac{1}{2} \nabla^a \Phi\left( 2\nabla^b h_{ab} -\nabla_a h  \right) 
\nonumber\\
-h^{ab}\nabla_a\nabla_b\Phi   + 2 h^{ab}\nabla_a \Phi\nabla_b \Phi=0\,,
\label{eq:flutuations2}
\end{align}
where the covariant derivatives and Riemann tensor refer to the background and $h=h_a^{\,a}$. 
Field perturbations will be classified according to the $SO(1,3)$ global symmetry
of the background. Thus we shall decompose the 
metric perturbations $h_{ab}$ as
\begin{align}
&h_{\alpha\beta}= h^{TT}_{\alpha\beta} + 
\partial_{(\alpha} h_{\beta)}^T + \left( 4\partial_\alpha\partial_\beta - \, \eta_{\alpha \beta} \partial^2 \right)  \bar{h} 
+  \eta_{\alpha\beta} h\,,
\nonumber\\
&
h_{zz}\,, \quad\quad\quad
h_{z\alpha}= v_{\alpha}^T +\partial_\alpha s\,.
\end{align}
As usual transverse and traceless (TT)
tensor fluctuations, transverse (T) vector fluctuations and scalar  fluctuations decouple.
Moreover, since we are interested only in the TT
metric fluctuations, we do not need to worry about mixing of perturbations. It is then simple to see that 
(\ref{eq:flutuations}) gives for the TT metric fluctuations
\begin{equation}
 \left( \nabla^2 
 -2e^{-2A(z)}  \dot{\Phi} \nabla_z 
 +2   \dot{A}^2 e^{-2A(z)}  \right) 
 h_{\alpha\beta}^{TT}=0\,.
 \label{eq:metric_fluctuations}
\end{equation}
The term with the dilaton arises from the usual coupling $-2\partial^c \Phi  \nabla_c h_{ab}$ for  metric fluctuations 
in the string frame; the other term comes from the coupling to the  Riemann tensor $R_{a c b d} h^{cd}$, with
${R_{\alpha \mu \beta \nu}} = \dot{A}^2 e^{2A}\left( \eta_{\alpha\nu} \eta_{\mu\beta} -  \eta_{\alpha\beta} \eta_{\mu\nu}   \right)$
and
${R_{\alpha z \beta z}} = -\ddot{A} e^{2A} \eta_{\alpha\beta}$.
In the case of pure AdS space, $A(z)= \ln (L/z)$, so (\ref{eq:metric_fluctuations}) simplifies to
\begin{equation}
\left(  \nabla^2 
 - m^2  \right)  h_{\alpha\beta} ^{TT}=0\,,
\label{eq:AdSgraviton}
\end{equation}
with $(Lm)^2=-2$, as expected  for the AdS graviton.
 
\subsection{\label{sec:SpinJ}Dual spin $J$ field}

\begin{figure}[t!]
\begin{center}
\includegraphics[height=6cm]{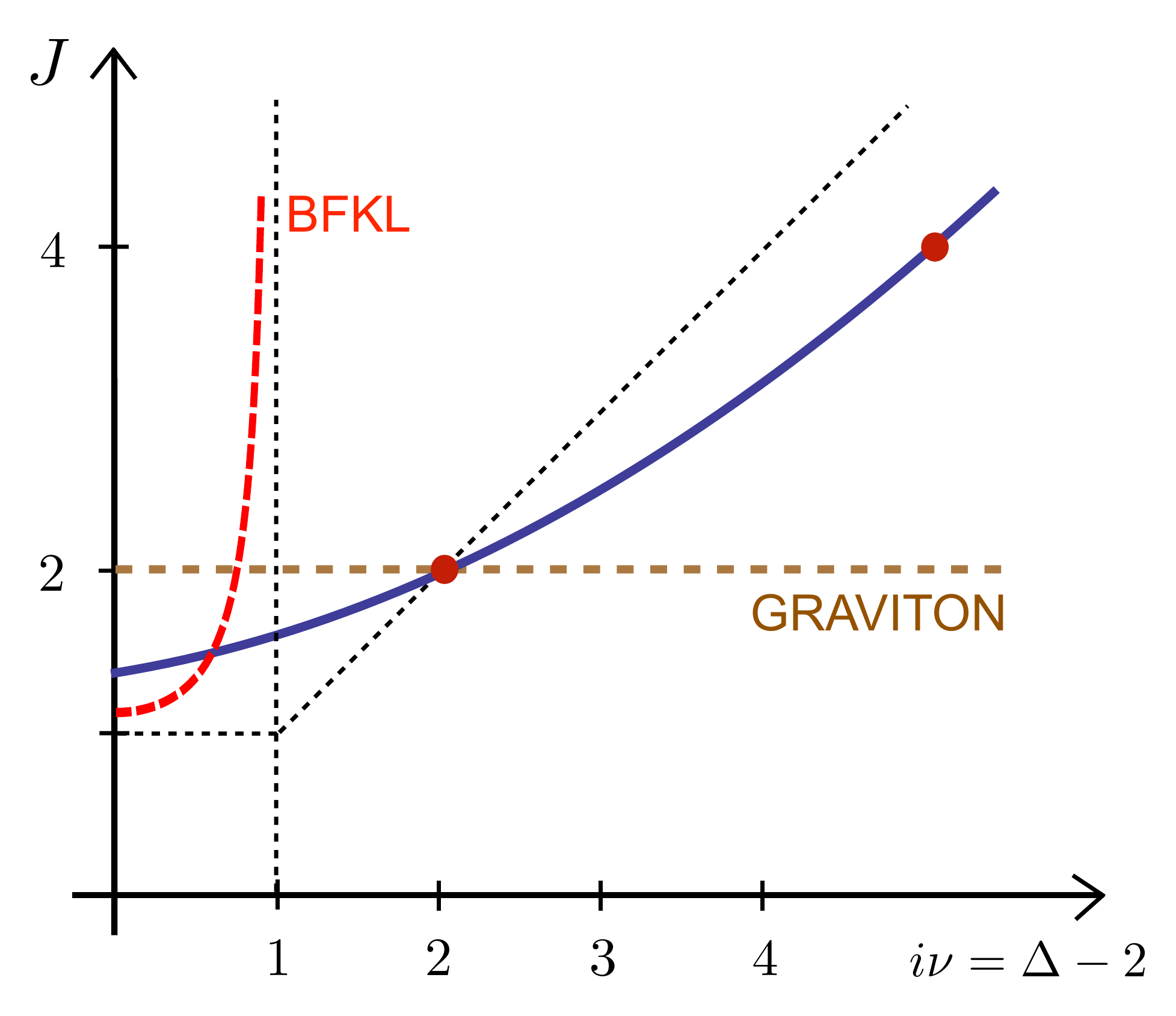}
\end{center}
\vspace{-0.5cm}
\caption{Expected form of the $\Delta=\Delta(J)$ curve (in blue).}
\label{fig:Delta(J)}
\end{figure}

We shall consider the exchange of  twist 2 operators of Lorentz spin $J$ formed from the gluon field \footnote{In the singlet sector there is also 
the twist 2 quark operator ${\cal O}_J \sim \bar{\psi} \Gamma_{\alpha_1} D_{\alpha_2} \cdots D_{\alpha_J} \psi$, however we are considering processes dominated by exchange of the gluon field.}
\begin{equation}
{\cal O}_J \sim {\rm tr} \left[ F_{\beta\alpha_1}  D_{\alpha_2} \cdots  D_{\alpha_{J-1}} F^{\beta}_{\ \alpha_J} \right] ,
\label{eq:twist2ops}
\end{equation}
where $D$ is the QCD covariant derivative. 
The dimension of the  operator ${\cal O}_J$ can be written as  $\Delta=2+J + \gamma_J$, where $\gamma_J$ is the anomalous dimension. 
In free theory the operator has critical dimension $\Delta=2+J$. 

Knowledge of the curve $\Delta=\Delta(J)$ is important 
when summing over spin $J$ exchanges, since this sum is done by analytic continuation in the $J$-plane, and then by considering the region of real $J<2$.
Figure \ref{fig:Delta(J)} summarizes a few important facts about the curve $\Delta=\Delta(J)$. Let us define the variable $\nu$ by $\Delta=2+i\nu$, and consider the inverse function $J=J(\nu)$. The figure shows the perturbative BFKL result for $J(\nu)$, which is an even function of $\nu$ and has poles at $i\nu=1$. 
Beyond perturbation theory, the curve must pass through the energy-momentum tensor
protected point at $J=2$ and $\Delta=4$. 
We shall use a quadratic approximation to this curve that passes through this protected point,
\begin{equation}
J(\nu)\approx J_0 - {\cal D} \nu^2\,,\ \ \ \ 4{\cal D}=2-J_0\,.
\label{eq:difusion}
\end{equation}
The use of a quadratic form for the function $J(\nu)$ is known as the diffusion limit and it is used both in
BFKL physics and in dual models that consider the AdS graviton Regge trajectory  
(see for instance \cite{Costa:2012fw}).

Consider now the spin $J$ field dual to the twist 2 operators (\ref{eq:twist2ops}). For pure AdS this field obeys the equation
\begin{equation}
 \left(  \nabla^2 
 - m^2  \right)  h_{a_1\cdots a_J} =0\,,\ \ \ 
 (Lm)^2=\Delta(\Delta-4) - J\,,
 \label{eq:AdS_SpinJ}
\end{equation}
where $L$ is the AdS length scale.
Note that this field is  symmetric, traceless 
and transverse
($ \nabla^b h_{b a_2\cdots a_J}=0$).

To  consider the spin $J$ field in a  general background of the form (\ref{eq:metric}), we need again to do a decomposition
in $SO(1,3)$ irreps. The propagating degrees are described by components $h_{\alpha_1\cdots \alpha_J}$, 
since the other components $h_{z\cdots z \alpha_i\cdots \alpha_J}$  ($i\ge 2$) are fixed 
by the transversality condition. Thus 
we need to define the equation of motion for $h_{\alpha_1\cdots \alpha_J}$. 
Of course we do not know its  form for the dual of QCD, but follow a phenomenological approach. We shall require that such
equation is compatible with the spin $2$ case (\ref{eq:metric_fluctuations}), since in that case it must reduce to that of the 
graviton, whose dual operator has protected dimension. Moreover, we require the coupling to the dilaton to be that of closed
strings in the graviton Regge trajectory arising from the term $-2\partial^c \Phi  \nabla_c h_{a_1\cdots a_J}$. Finally,
we require the equation to reduce to (\ref{eq:AdS_SpinJ}) in the conformal limit (constant dilaton).
This leads to the following proposal
\begin{equation}
\left( \!  \nabla^2 
\!-\!2 \,e^{-2A} \dot{\Phi}  \nabla_z \!-\frac{\Delta (\Delta -4)}{L^2}  +\! J \dot{A}^2 e^{-2A} \!\right) \!h_{\alpha_1 \dots \alpha_J}\!=0\,,
\label{eq:dualQCD_SpinJ}
\end{equation}
where here $L$ is a length scale parameter. It is trivial to verify that setting  $J=2$ (and $\Delta=4$) this equation reduces to the the graviton
equation (\ref{eq:metric_fluctuations}). Similarly, setting $A(z)= \ln (L/z)$ and $\Phi=0$ we recover the spin $J$ AdS equation (\ref{eq:AdS_SpinJ}).
The dilaton term arises from considering tree level closed strings, which is justified since we work at large $N$. 
We expect that there will be more terms in this equation arising from other curvature couplings and derivatives of the dilaton field. 
Assuming the equation is analytic in $J$, these terms should be proportional to 
$J-2$, so that they are absent for $J=2$.  Notice that there can be such terms still at the level of two derivatives, that is terms proportional to
\begin{equation}
e^{-2A} \left(\dot{A}^2 - \ddot{A}\right) ,\quad\quad
e^{-2A}\dot{\Phi}^2\,,\quad\quad
e^{-2A}\ddot{\Phi}\,,
\end{equation}
which also vanish in the conformal limit. Terms with higher derivatives will appear in a $\alpha'/L^2$ expansion.
 As already stated, we shall follow a phenomenological approach
and use the simple form (\ref{eq:dualQCD_SpinJ}) to describe the fluctuations of the spin $J$ field  in holographic QCD.


We will be interested in the continuation of  (\ref{eq:dualQCD_SpinJ}) to the unphysical
region of $J<2$. It is here that we will use the  diffusion limit (\ref{eq:difusion}), writing in (\ref{eq:dualQCD_SpinJ})
\begin{equation}
\frac{\Delta (\Delta -4)}{L^2}  \approx \frac{2}{l_s^2}\, (J-2)\,,
\label{eq:difusion2}
\end{equation}
with $l_s$ a length scale set by the QCD string. Notice that we are fixing $l_s$ to a constant determined by IR physics, but in fact 
it should depend on energy scale, since the curve $\Delta=\Delta(J)$ in Figure \ref{fig:Delta(J)} should vary with energy scale, keeping its general shape. 
However, for the soft-pomeron   this should not matter \footnote{For instance, the approximation (\ref{eq:difusion2})  misses 
the dimensions of the operators with $J>2$ in the free theory limit.}. We leave $l_s$ as a phenomenological parameter to be fixed by data.

In the Regge limit we are actually interested in the $+\cdots+$ component of (\ref{eq:dualQCD_SpinJ}). 
To find the solution write
\begin{equation}
h_{+\cdots+} (z,x)= e^{i q\cdot x} e^{\frac{2J-3}{2}A(z) +\Phi(z)} \psi(z)\,,
\end{equation}
where $q\cdot x=\eta_{\alpha\beta}q^\alpha x^\beta$ and we set  $q_-=0$ in the Regge limit. Then, a  computation shows that
(\ref{eq:dualQCD_SpinJ}) reduces to the Schr\"{o}dinger problem
\begin{align}
&\left( -\frac{d^2\  }{dz^2} + U(z) \right) \psi(z) = t \,\psi(z)\,,
\label{eq:Schrodinger}
\\
&U(z) = \frac{15}{4}\,\dot{A}^2 - 5\dot{A}\dot{\Phi} + \dot{\Phi}^2 + 
\frac{\Delta (\Delta -4)}{L^2} \,e^{2A(z)}\,,
\label{eq:Potential_U}
\end{align}
with $t=-q^2$.
The energy spectrum for each $J$  quantises $t=t_n(J)$, therefore yielding the
glueball  masses.

\subsection{\label{sec:SpinJexchange}t-channel spin J exchange}

Next consider the elastic scattering of QCD hadronic states of masses $m_1$ and $m_2$.
We write the incoming momenta $k_1$, $k_2$  and the outgoing momenta $k_3$,
$k_4$ in  light-cone coordinates $(+,-,\perp)$ as
\begin{align}
&k_1=\left(\!\sqrt{s},\frac{m_1^2}{\sqrt{s}} ,0\right),\  \ k_3=-\left(\!\sqrt{s},\frac{ m_1^2+ q_\perp^2}{\sqrt{s}} , q_\perp \right)\!,\\
&k_2=\left(\frac{m_2^2}{\sqrt{s}},\sqrt{s} ,0\right),\  \ k_4=-\left(\frac{m_2^2+ q_\perp^2}{\sqrt{s}},\sqrt{s} ,-q_\perp \right),
\nonumber
\end{align}
where we consider the  Regge limit $s\gg t=-q_\perp^2$.

Each hadron is described by a normalizable mode $\Upsilon_i(z,x) = e^{ik_i\cdot x_i}\upsilon_i(z)$
where $\upsilon_3=\upsilon_1^*$ and $\upsilon_4=\upsilon_2^*$.
The hadrons we consider are made of open strings. Then the coupling of each hadronic field to the 
spin $J$ closed string fields has the form
\begin{equation}
\kappa_J \int d^5x \sqrt{-g}\, e^{-\Phi}   h_{a_1\cdots a_J}\Upsilon\nabla^{a_1}\cdots \nabla^{a_J} \Upsilon\,.
\end{equation}
Notice that in principle different types of hadrons will have a different coupling $\kappa_J$.
The transverse condition on the spin $J$ field guarantees that this coupling is unique up to  derivatives of the
dilaton field, which are subleading in the Regge limit.

The amplitude for $m_1m_2\to m_1m_2$  scattering through  exchange of a spin $J$ field in the t-channel
may now be computed  in the dual theory. In the Regge limit we have
\begin{align}
{\cal A}_J(k_i) &= - \kappa_J \kappa'_J \int d^5Xd^5X'  \sqrt{-g} \sqrt{-g'}  e^{-\Phi-\Phi'}
\nonumber
\\
&\left(\Upsilon_1\partial_-^J\Upsilon_3  \right)\Pi^{-\cdots-,+\cdots+}(X,X') \big(\Upsilon'_2\partial'_+{\!\!^{J}}\Upsilon'_4\big)\,,
\label{eq:spinJ_exchange}
\end{align}
where $X=(z,x)$ and $X'=(z',x')$ are  bulk points and fields with a prime are evaluated at $X'$,
e.g. $\Phi'\equiv\Phi(z')$. We use this notation throughout. We expect the spin $J$ field propagator
to obey an equation of the type
\begin{align}
&( {\cal D} \Pi)_{a_1\cdots a_J,b_1\cdots b_J}(X,X') =
\nonumber\\
& i e^{2\Phi}g_{a_1\left(b_1\right.} \!\cdots g_{|a_J| \left. b_J\right)} \delta_5(X,X') - {\rm traces}\,,
\label{eq:SpinJpropagator_eq}
\end{align}
for some differential operator ${\cal D}$.
We are interested in the ${+\cdots+,-\cdots-}$ component  of this equation, 
for which the differential operator ${\cal D}$ can be read from (\ref{eq:dualQCD_SpinJ}).

Some algebra shows the amplitude (\ref{eq:spinJ_exchange}) simplifies to
\begin{align}
{\cal A}_J(s,t) =&\, i V \frac{\kappa_J \kappa'_J}{ (-2)^{J}} \,s \int dz dz'  e^{3A+3A'-\Phi-\Phi'}
\nonumber\\
&|\upsilon_1|^2 |\upsilon'_2|^2 \left( s e^{-A-A'}\right)^{J-1} G_J(z,z',t)\,,
\end{align}
where $V$ is the boundary volume.  The function
\begin{equation}
G_J(z,z',t) = \int d^2l_\perp e^{-iq_\perp\cdot l_\perp} G_J(z,z',l_\perp)\,,
\end{equation}
is the Fourier transform of
\begin{align}
G_J(z,z',l_\perp)=&\,i(-2)^J e^{(1-J) (A+A')}  
\nonumber
\\
&\frac{1}{2}\int dw^+dw^- \Pi_{+\cdots+,-\cdots-}(z,z',w)\,,
\end{align}
where $w=x-x'=(w^+,w^-,l_\perp)$ and $l_\perp=x_\perp-x'_\perp$.
From the ${+\cdots+,-\cdots-}$ component of (\ref{eq:SpinJpropagator_eq}), 
as defined by (\ref{eq:dualQCD_SpinJ}), it follows that
$G_J(z,z',l_\perp)$ is an Euclidean scalar propagator in the three-dimensional transverse space
 of the dual scattering process ($dx^+=dx^-=0$ in (\ref{eq:metric})), i.e.
\begin{align}
&\bigg[ \Box_3 - 2 e^{-2A(z)} \dot{\Phi} \partial_z - e^{-2A(z)} \left( 2\dot{A}^2 + \ddot{A} -2\dot{A}\dot{\Phi}  \right) 
\nonumber
\\
&\left.
  - \frac{\Delta(\Delta-4)}{L}\right]G_J(z,z',l_\perp)=- e^{2\Phi} \delta_3(x,x')\,,
  \label{eq:Gprop_eq}
\end{align}
where here $x=(z,x_\perp)$ and $x'=(z',x'_\perp)$. 
Writing
\begin{equation}
G_J(z,z',t)= e^{\Phi(z)-\frac{A(z)}{2}} \psi(z)\,,
\end{equation}
the homogeneous solution to (\ref{eq:Gprop_eq}) is exactly given by the Schrodinger problem 
of (\ref{eq:Schrodinger}) and (\ref{eq:Potential_U}). 
Moreover, using $\sum_n \psi_n(z)\psi^*_n(z')=\delta(z-z')$, 
we conclude that
\begin{equation}
G_J(z,z',t)= e^{\Phi-\frac{A}{2}+\Phi'-\frac{A'}{2}} \sum_n  \frac{\psi_n(z)\psi^*_n(z')}{t_n(J)-t}\,.
\end{equation}
Note  the eigenvalues $t_n$ and functions $\psi_n$ depend on $J$.

\subsection{\label{sec:Regge}Regge theory}

\begin{figure}[t!]
\begin{center}
\includegraphics[height=6cm]{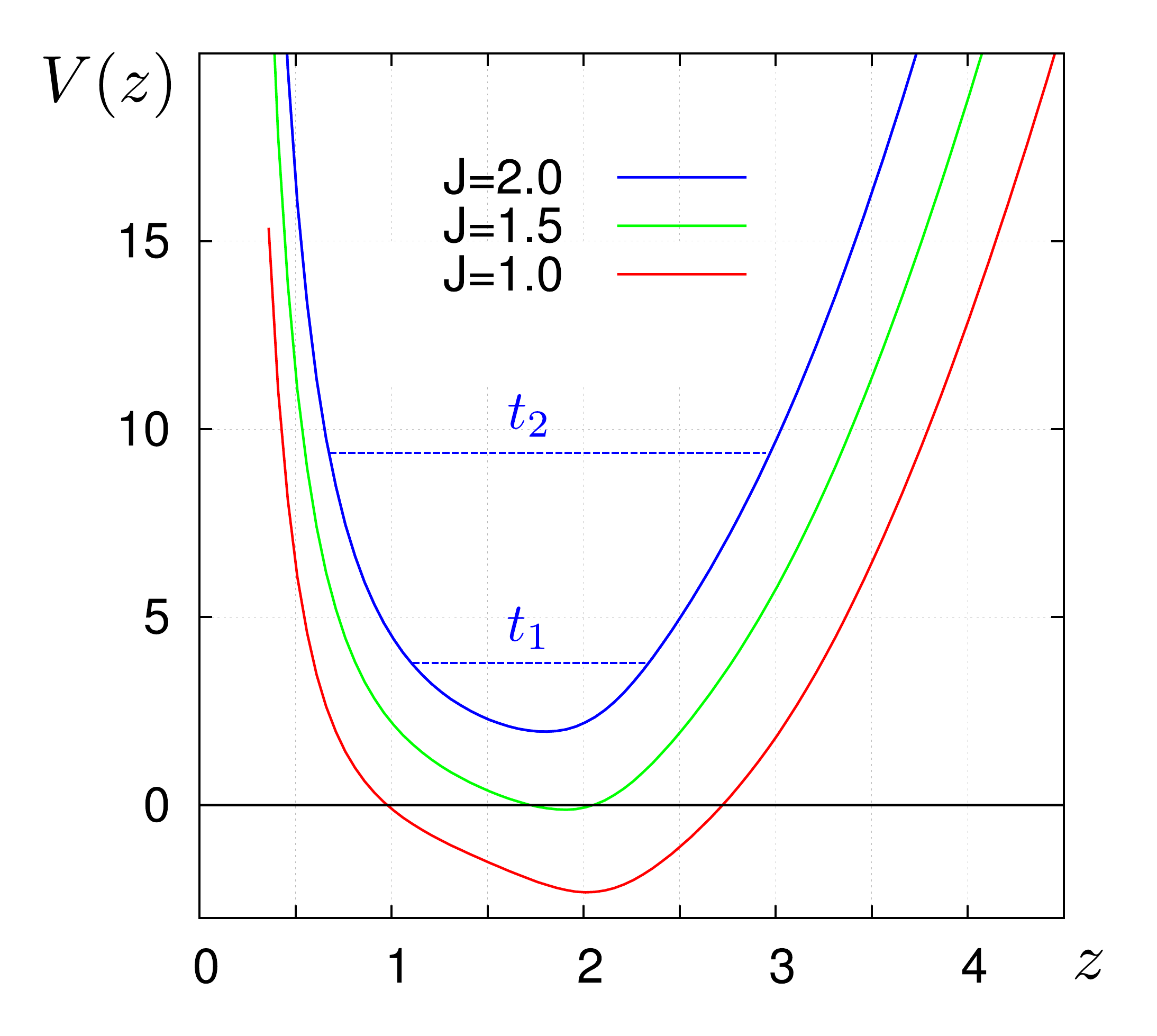}
\end{center}
\vspace{-0.5cm}
\caption{Effective potential for different values of spin $J$. The first $2^{++}$ glueball
states are also shown.}
\label{fig:Potential}
\end{figure}

We will sum all even spin $J$ exchanges with $J\ge 2$ using
a Sommerfeld-Watson transform
\begin{equation}
\frac{1}{2} \sum_{J\ge 2} \left(  s^J + (-s)^J\right) \to
-\frac{\pi}{2} \int \frac{dJ}{2\pi i} \frac{s^J + (-s)^J}{\sin(\pi J)}\,,
\end{equation}
which requires the analytic continuation of the amplitude 
${\cal A}_J(s,t)$ to the complex $J$-plane.
Then, the  amplitude for the exchange 
off all even spin $J$ fields becomes
\begin{equation}
{\cal A}(s,t) = iV  \int dz dz' e^{3(A+A')}| \upsilon_1|^2 |\upsilon'_2|^2 \sum_n \chi_n\,,
\label{eq:Amplitude}
\end{equation}
where $\chi_n=\chi_n(z,z',s,t)$ is given by
\begin{align}
\chi_n=& -\frac{\pi}{2} \int \frac{dJ}{2\pi i}   \frac{s^J + (-s)^J}{\sin(\pi J)} \frac{\kappa_J \kappa'_J}{2^J}
\nonumber\\
& e^{-(J-\frac{1}{2})(A+A')}  \frac{\psi_n(z)\psi^*_n(z')}{t_n(J)-t}\,.
\end{align}
We assume the $J$-plane integral can be deformed from the 
poles at even values of $J$, to the poles  $J=j_n(t)$ defined by $t_n(J)=t$. 
In the  scattering domain of negative $t$ these poles are  along the real axis for $J<2$.
Thus we can write
\begin{align}
\chi_n &= s^{j_n(t)} \Big [-\frac{\pi}{2} \left(  \cot \frac{\pi j_n}{2} + i \right)
  \frac{\kappa_{j_n} \kappa'_{j_n}}{2^{j_n}} 
  \nonumber\\
&e^{-(j_n-\frac{1}{2})(A+A')}  \frac{d j_n}{dt} \psi_n(z)\psi^*_n(z') \Big ] \,,
\end{align}
where $j_n=j_n(t)$ and we remark that the wave functions $\psi_n$ are computed at
$J=j_n(t)$. It is clear that for large $s$ the amplitude (\ref{eq:Amplitude}) will be dominated 
by the Regge pole with highest $j_n(t)$, in accord with the Regge 
behavior (\ref{eq:SoftPomeron}).  

\begin{figure}[t!]
\begin{center}
\includegraphics[height=6cm]{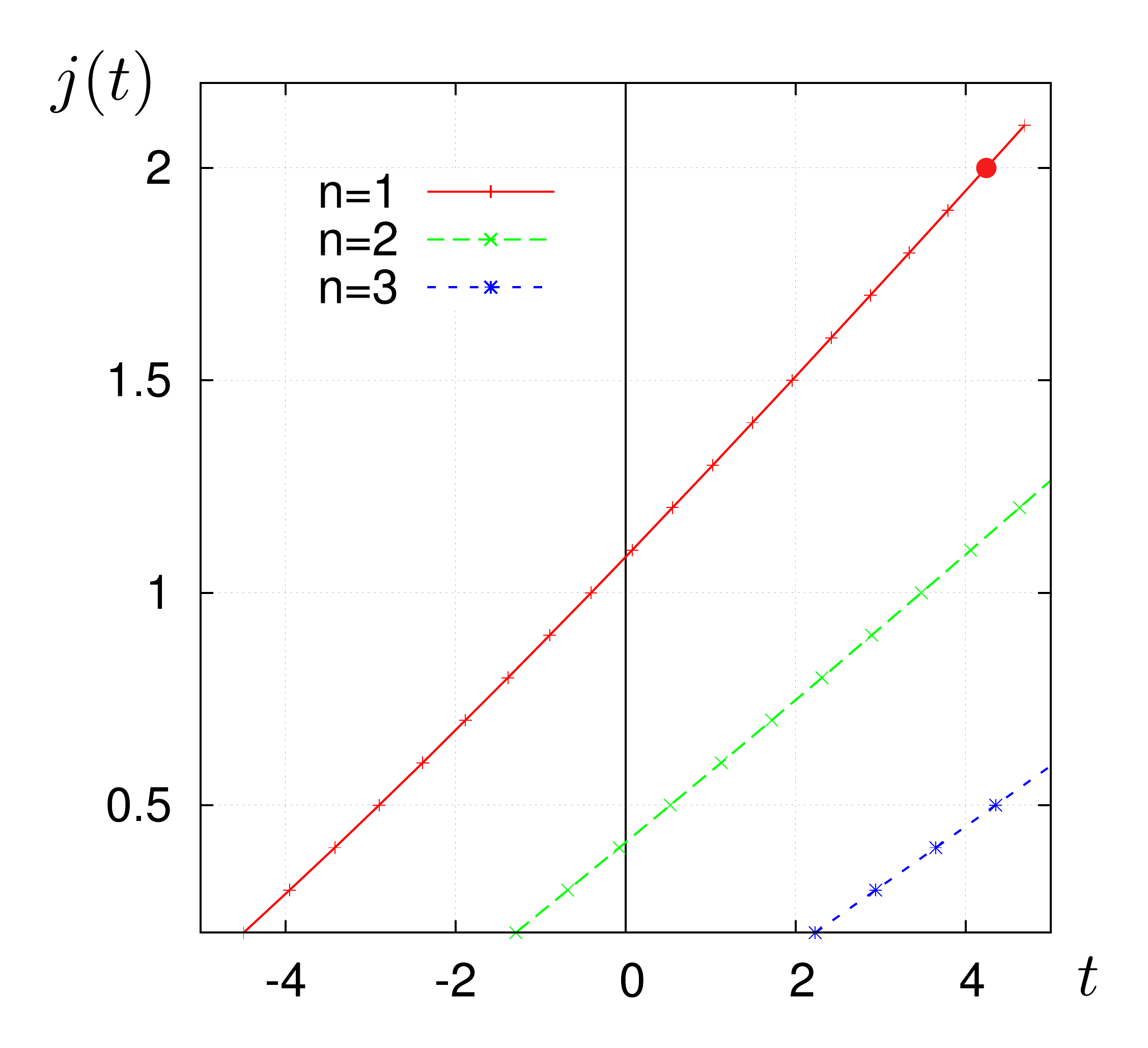}
\end{center}
\vspace{-0.5cm}
\caption{The first Regge trajectories that result from solving the Schr\"{o}dinger problem for discrete values of $J$.
}
\label{fig:ReggeTrajectories}
\end{figure}

We now specify to the model 
considered in this letter, which is determined by the effective Schr\"{o}dinger 
potential (\ref{eq:Potential_U}). Since we are interested in the region $J<2$, we can use the 
model introduced in (\ref{eq:difusion2}) for the curve $\Delta=\Delta(J)$. Figure \ref{fig:Potential}
shows the potential for several value of $J$. The energy levels for $J=2$ are shown and 
compute the mass of the spin 2 glueball masses. As $J$ decreases the  energy levels will eventually 
cross the zero energy value. This will be the value of the intercept for the $n$-th Reggeon. 
Figure \ref{fig:ReggeTrajectories} shows the curves $j_n(t)$, which clearly show that $n=1$ is the leading Regge pole.
The curves are approximately straight so we can also define a Regge slope. Note that the model considered in this paper allows us to investigate the region of real $J<2$ and find the Regge poles. This differs from previous approaches based on Regge trajectories (see for instance \cite{BoschiFilho:2005yh}).

\section{\label{sec:Results}Results}

Finally we can  test  to  which degree we are reproducing QCD physics. 
We consider first the leading Regge pole.  
We vary $l_s$, introduced in (\ref{eq:difusion2}), to fix the Pomeron intercept to the value given in 
\cite{Donnachie:1992ny}, as an optimal fit for total cross sections. 
For the  value $l_s=0.178 \ {\rm GeV}^{-1}$, and independently of our choice of $\Lambda_{QCD}$,
we obtained  $\alpha_0= 1.08$.  The  value of the slope is then fixed by the choice of $\Lambda_{QCD}$. We obtained
$\alpha' \Lambda_{QCD}^2=0.018$.
If we fix $\Lambda_{QCD}=0.292\ {\rm GeV}$ as in \cite{Gursoy:2007cb,Gursoy:2007er}, 
such that the first glueball mass $m_{0^{++}}=1.475\ {\rm GeV}$, one obtains $\alpha'=0.21 \ {\rm GeV}^{-2}$. 
If, on the other hand, we require the measured value of $\alpha'=0.25 \ {\rm GeV}^{-2}$ \cite{Jaroszkiewicz:1974ep}, we obtain 
$\Lambda_{QCD}=0.265$. This is consistent with having the $2^{++}$  glueball of the Pomeron trajectory with a mass of $1.9\ {\rm GeV}$,
which is a known possibility \cite{DL:Book,Brunner:2015oqa}. 

Let us  remark that we could fix $l_s$ to reproduce the intercept obtained in lattice simulations of $SU(3)$
pure Yang-Mills \cite{Meyer:2004jc}. In this case, for $l_s = 0.192 \ {\rm GeV}^{-1}$ one has $\alpha_0= 0.93$. Then, setting
$\Lambda = 0.292$, which is fixed to reproduce $m_{0^{++}}=1.475\ {\rm GeV}$ of the same lattice simulations, we obtained 
a slope $\alpha'=0.25 \ {\rm GeV}^{-2}$. This is exactly the slope obtained by the lattice simulations \cite{Meyer:2004jc}.

\begin{figure}[t!]
\begin{center}
\includegraphics[height=6cm]{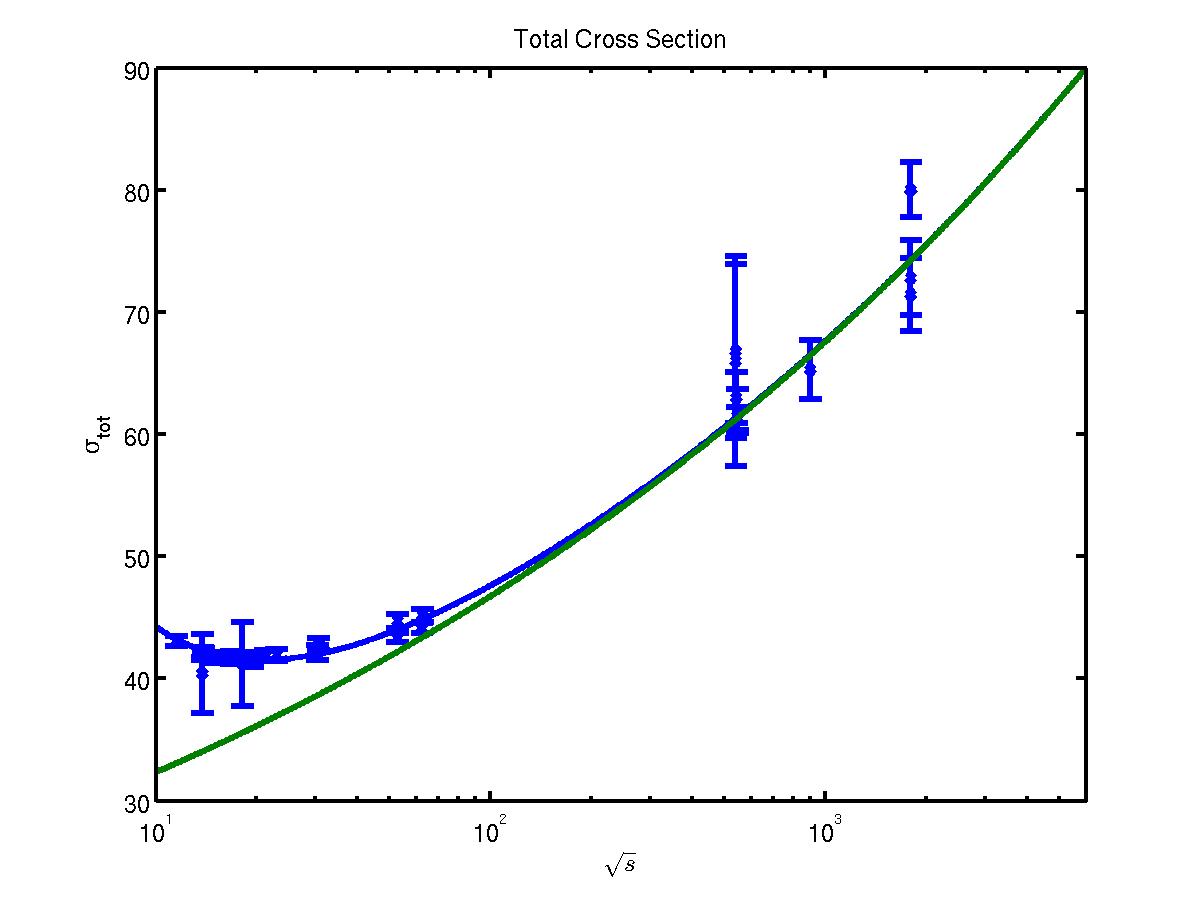}
\end{center}
\vspace{-0.5cm}
\caption{A fit to $p\,\bar{p}$ total cross section data using the exchange of the first two Regge poles in our model. The green line represents the leading Pomeron exchange, and fails to fit the data at moderate values of $\sqrt{s}$.}
\label{fig:cross_section}
\end{figure}

For the second pole we obtained an intercept  of $0.433$, which is consistent with the value used in \cite{Donnachie:1992ny}.
We ran fits to $p\,\bar{p}$ total cross section data \cite{data}  and found that the second pole is necessary and needs to be in a narrow range of $\approx 0.35 - 0.55$. We determined this range by fitting an expression of the form
\begin{equation}
\sigma = g_0 (\alpha' s)^{\alpha_0} + g_1 (\alpha' s)^{\alpha_1}\, ,
\end{equation}
using $g_0$ and $g_1$ as parameters, and varying $\alpha_1$. We fit this to $p\, \bar{p}$ scattering data with $\sqrt{s} > 10 \, \text{GeV}$. The above range is fixed by the requirement that $\chi^2_{d.o.f.}$ be of order 1 or less. Our results can be seen in figure \ref{fig:cross_section}.
As can be seen there using just the leading Pomeron exchange fails to fit the data satisfactorily.
The second pole in \cite{Donnachie:1992ny} corresponds to several degenerate meson trajectories, while here it  represents a next-to-leading glueball trajectory. Thus, our work points to the possibility that 
in this range there is a glueball trajectory as well. In fact, at least some of the $f_2$ states are known to correspond to glueballs (see \cite{Sonnenschein:2015zaa} and references therein for recent results).

\section{\label{sec:Conclusion}Conclusion}

Soft pomeron physics is still beyond the current analytic understanding of QCD. The best one can do at weak coupling is to start from the BFKL approach
and then introduce the running of the coupling, therefore breaking conformal symmetry. As  a consequence, the branch cut of the BFKL pomeron becomes a set
of poles in the $J$-plane \cite{Kowalski:2010ue}. This approach can be used to fit DIS data for hard scattering, keeping a very large number of poles. However, it is not applicable to the case of soft probes. 
In general we expect to have a description of soft pomeron exchange as a Regge pole, in agreement with the phenomenological approach pioneered by Donnachie and Landshoff \cite{Donnachie:1992ny}.
Such description was proposed in \cite{Brower1}, based on scattering of closed strings in a  dual confining background. In particular that work anticipated that for confining theories with  a negative $\beta$ function 
the pomeron, described as the graviton Regge trajectory, becomes a Regge pole. Our work confirms this expectation by  extending 
the holographic QCD model of \cite{Gursoy:2010fj} to scattering processes dominated by soft pomeron exchange, bringing a new insight to soft pomeron physics.

 Let us finish with a caveat and two open questions. 
It has been claimed that a soft Pomeron pole is not enough to describe the new 
 LHC data \cite{newfits}. This is somewhat expected, since it is known that at very high energies such a Regge pole would violate the Froissart-Martin bound, and other effects need to be included,  for example multi-Pomeron exchange. 
 However, this does not invalidate the great experimental successes of soft Pomeron exchange up to LHC energies, as well as 
 the necessity to understand the subleading trajectories.

The first question concerns the relation between hard and soft pomerons. 
Recent studies in gauge/gravity duality  reproduce a plethora of low-$x$ processes using the graviton Regge trajectory 
as the dual trajectory of  the QCD Pomeron \cite{Saturation,Levin:2010gc,Brower:2010wf,Costa:2012fw,Costa:2013uia,Brower:2012mk}. 
In these cases  one observes  a running of the intercept with the  size of the probes.  It would be very interesting if we could embed these results within the present model, therefore unifying 
both pomerons. Another question is related to the spectrum of the spin $J$ field, at integer values. It would be very nice to reconstruct the 
spin $J$ equation in this domain, such that it reproduces perturbative QCD results.




\begin{acknowledgments}
We wish to thank Nick Evans, Jo\~ao Penedones, Chung-I Tan and Dimitrios Zoakos for  discussions. 
This research received funding from the [European Union] 7th Framework Programme (Marie Curie Actions) under grant agreements No 269217 and 317089 (GATIS).
The work of A.B. has been supported by the Brazilian agency CAPES, through the fellowship BEX 8051/14-3. A.B also thanks  the Galileo Galilei Institute for Theoretical Physics for the hospitality and the INFN for partial support during the completion of this work.
The work of M.D. has been supported by the Portuguese Funda\c{c}\~ao para a Ciencia e a Tecnologia (FCT) through the fellowship SFRH/BCC/105757/2014.
\end{acknowledgments}

\end{document}